\documentclass[iop,apj,numberedappendix]{emulateapj}
\usepackage{amsmath,amssymb,amstext}
\usepackage[breaklinks,colorlinks,citecolor=blue,linkcolor=magenta]{hyperref}

\usepackage[all]{hypcap} 
\usepackage{aas_macros}
\usepackage{natbib}
\shorttitle{Nebular Spectra of SNe~Ia with Stripped Companion Gas}
\shortauthors{Boty\'{a}nszki et al.}

\begin{document}

\newcommand{\fsix}[1]{$^{56}$#1}
\newcommand{\Msun}[0]{$\rm M_\odot$}
\newcommand{\ha}[0]{H$\alpha$}
\newcommand{\sub}[2]{${#1}\rm _{#2}$}
\newcommand\T{\rule{0pt}{2.6ex}}       
\newcommand\B{\rule[-1.2ex]{0pt}{0pt}} 

\title{Multidimensional Models of Type Ia Supernova Nebular Spectra: Strong Emission Lines from Stripped Companion Gas Rule Out Classic Single Degenerate Systems}
\author{J\'{a}nos Boty\'{a}nszki\altaffilmark{1}, Daniel Kasen\altaffilmark{1,2,3}, and Tomasz Plewa\altaffilmark{4}}
\affil{\altaffilmark{1}Physics Department, University of California, Berkeley, CA 94720, USA}
\affil{\altaffilmark{2}Astronomy Department and Theoretical Astrophysics Center, University of California, Berkeley, CA 94720, USA}
\affil{\altaffilmark{3}Nuclear Science Division, Lawrence Berkeley National Laboratory, Berkeley, CA 94720, USA}
\affil{\altaffilmark{4}Department of Scientific Computing, Florida State University, Tallahassee, FL 32306, USA}

\begin{abstract}

The classic single-degenerate model for the progenitors of Type Ia Supernova (SN Ia) predicts that the supernova ejecta should be enriched with solar-like abundance material stripped from the companion star. Spectroscopic observations of normal SNe~Ia at late times, however, have not resulted in definite detection of hydrogen. In this Letter, we study line formation in SNe Ia at nebular times using non-LTE spectral modeling. We present, for the first time, multidimensional radiative transfer calculations of SNe Ia with stripped material mixed in the ejecta core, based on hydrodynamical simulations of ejecta-companion interaction. We find that interaction models with main sequence companions produce significant \ha\ emission at late times, ruling out this type of binaries being viable progenitors of SNe Ia. We also predict significant He~I line emission at optical and near-infrared wavelengths for both hydrogen-rich or helium-rich material, providing an additional observational probe of stripped ejecta. We produce models with reduced stripped masses and find a more stringent mass limit of $M_{\rm st} \lesssim 1\times 10^{-4} M_\odot$ of stripped companion material for SN 2011fe.

\end{abstract}

\maketitle

\section{Introduction}

While it is believed that Type Ia Supernovae (SNe~Ia) are the thermonuclear explosion of a carbon-oxygen white dwarf (WD) in a binary system, the nature of the companion star and the mechanism that triggers the disruption remain uncertain. SNe~Ia have been used successfully to infer the accelerating expansion of the universe \citep{riess1998, perlmutter1999}, and a better SN~Ia progenitors may lead to an improved ability to standardize these events. Various theoretical models can reproduce the basic observational properties of SNe~Ia \citep{wang2012, hillebrandt2013}, but work is needed to determine which models explain which specific subsets of observed SNe Ia. 

SN Ia models fall into either the single degenerate (SD) or double degenerate (DD) category, depending on the nature of the WD's binary companion. A key prediction of SD progenitor models is that material from companion star will be swept up by SN~Ia ejecta \citep{wheeler1975, chugai1986} and may be detectable in late-time spectra. Hydrodynamical simulations have found that the mass of stripped companion material is typically between 0.002 and 0.5\Msun, depending on the type of companion and binary system properties \citep{marietta2000, meng2007, pakmor2008, pan2010, pan2012, liu2012, liu2013a_HeCompanions, boehner2017}. This material is embedded in the ejecta at low velocities ($\approx$ 1000 -- 2000 km s$^{-1}$) and may be observable as narrow emission lines at late times ($\gtrsim 200$~days) when the ejecta become optically thin. 

Many observations at nebular times have failed to detect Balmer emission (specifically, \ha) in late-time spectra \citep{mattila2005, leonard2007, shappee2013, bikmaev2015, lundqvist2013, lundqvist2015, maguire2016}. The only known exception may be the 3.1$\sigma$ detection of \ha\ in SN~2013ct by \citet{maguire2016}. Translating the flux limits into a reliable constraint on the stripped mass, however, requires multidimensional hydrodynamic and radiative transfer calculations which have not, to date, been conducted. 

Most analyses of late-time SN~Ia spectra derive mass constraints from 
\citet{mattila2005}, who studied \ha\ formation at late times using parameterized spherically symmetric radiative transfer calculations. 
These models assumed the W7 SN ejecta profile \citep{nomoto1984, thielemann1986} and added by hand varying amounts of uniform density solar-abundance material at the center of the ejecta up to a fixed velocity of 1000 km~s$^{-1}$. These authors concluded that stripped masses $\gtrsim 0.03$~\Msun\ should produce detectable Balmer lines in the nebular spectra of SNe Ia, in conflict with observations.

Hydrodynamical models of the interaction of SNe Ia with a companion star, however, find that the stripped companion material strongly varies in density and its distribution is highly non-uniform. How this affects the predicted emission line strengths is unclear, and requires multidimensional modeling. 

In this Letter, we study realistic distributions of stripped companion material derived from multi-D hydrodynamical simulations which we post-process with a multi-D non-local thermodynamic equilibrium (NLTE) radiative transport code to synthesize nebular spectra as a function of viewing angle. This allows us to predict the line strengths resulting from stripped material of different total mass and composition, and from a variety of different companion stars (main sequence stars, subgiants, and red giants).

\section{Methods}
\label{sec:methods}

The SN Ia ejecta-companion interaction models used in this work are taken from \citet{boehner2017}, henceforth B17. Assuming cylindrical symmetry, they obtained well-resolved hydrodynamic models of interaction for a sample of semi-detached binaries considered as potential SN~Ia SD progenitor channels. The sample includes systems with various main sequence (MS), subgiant (SG), and red giant (RG) companions. The spherically symmetric W7 model \citep{nomoto1984, thielemann1986} is used as the explosion model for the supernova. The material stripped from the companion has solar metallicity, and we consider cooling from important metals like S, Si, Ca, and Fe.

We calculate nebular spectra using the 3D NLTE radiation transport code presented in \citet{botyanszki2017}, which takes as input a homologously expanding supernova model and calculates ejecta temperature and line emissivities based on the balance of radioactive heating and line emission. Energy deposition from radioactively produced gamma-rays is calculated using a Monte Carlo code \citet{kasen2006} while positrons are assumed to be deposited locally. The ejecta are assumed to be optically thin, though we adjust the radiative decay rates of optically thick lines using the formulation of \citet{liMcCray1993}. 

Recombinations are assumed to be transitions between ground states only, except in the case of hydrogen and helium, where recombination to excited states is expected to contribute non-trivially to optical line strengths. For these species, we use a simplified approach in which the recombination rates split evenly among levels and the sum of these is consistent with the well-known total recombination rates. While this recombination treatment can be improved by using level dependent values for the recombination rates, the ionization fractions of elements depend only on total recombination rates. Allowing recombination to excited states in hydrogen lowers optical line strengths, suggesting that collisional excitation from the ground state is the dominant way of populating excited states. On the contrary, allowing recombination to excited states in helium increases optical line strengths, suggesting that recombination contributes significantly to the population of excited states. The excitation of permitted helium transitions is therefore a result of both collisional excitation due to thermal electrons as well as recombination to excited states following collisional ionization due to non-thermal electrons.

\section{Results}
\label{sec:multiD}

\begin{figure*}[htbp]
\label{fig:fig1}
\includegraphics[width=\textwidth]{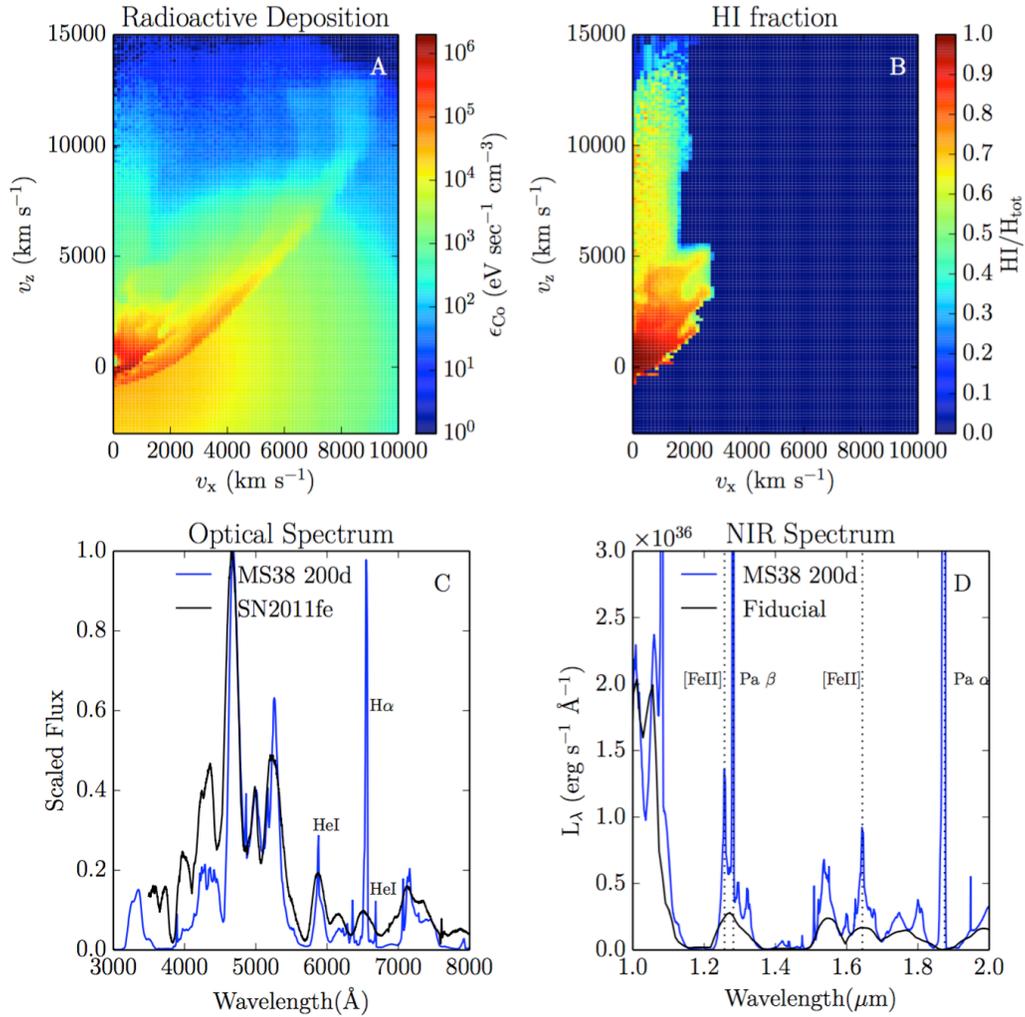}
\caption{Properties of the MS38 model at 200 days past explosion, zoomed in to show the region of mass stripped from the companion. Panel (A) shows the deposition of energy from \fsix{Co} decay throughout the ejecta, strongest in the region of stripped material at low velocities into which gamma-rays have penetrated from surrounding \fsix{Co}-rich regions. Panel (B) shows the calculated fraction of neutral hydrogen to total hydrogen. Dark blue regions contain less than 1\% H by mass. The region near the core has high density, thus strong recombination. Panel (C) shows the synthetic optical spectrum of the MS38 model. \ha\ is prominent next to the normal forbidden iron and cobalt features, and permitted He I lines are also visible. Panel (D) shows the synthetic NIR spectrum of the MS38 model. The fiducial model from \citet{botyanszki2017} is included in black for reference. Dotted lines identify the following transitions: [Fe II] 12.257$\mu$m, H I 1.2820$\mu$m (P$\beta$), [Fe II] 1.6440 $\mu$m, HI 1.8750$\mu$m (P$\alpha$). He I emission is also prominent at 1.085$\mu$m. Line profiles in the MS38 model are more peaked than in the fiducial model, and narrow hydrogen Paschen emission dominates the NIR spectrum.}
\end{figure*}

Panels A and B of Figure \ref{fig:fig1} show the structure of a sample model (MS38) from B17 at 200 days past explosion, zoomed in to show the hydrogen-rich region. The stripped material occupies a conical section of the remnant from which the nickel-rich ejecta have been pushed away. There is no mixing of the stripped material with the nickel-rich ejecta except near the interface. Gamma-rays from \fsix{Co} decay enter the hydrogen-rich region and deposit their energy there. Due to high densities in this region, recombination is efficient against non-thermal collisional ionization and, consequently, hydrogen stays in a mostly neutral state. 

Panels C and D of Figure \ref{fig:fig1} show the MS38-based synthetic nebular spectrum at 200 days at optical and NIR wavelengths, respectively. At that time, the peak \ha\ flux is very strong,  comparable to the prominent [Fe III] $\lambda$ 4658 emission line from the SN ejecta. In addition, narrow features from He I and [Fe II] are visible in the optical. In the near-infrared, lines from the Hydrogen Paschen series are conspicuous, along with narrow emission from forbidden iron group lines.

The luminosity of our calculated \ha\ emission is higher than that of previous parameterized 1D models \citep{mattila2005, lundqvist2013}. We attribute this to the more realistic geometry in our multi-D calculations. In the previous works, stripped material was assumed to occupy a spherically symmetric central region of constant density. However, the hydrodynamical models show that stripped material is concentrated in a smaller volume clump that is offset from the ejecta center. The density of the hydrogen-rich material is higher, leading to a more effectively trapping of gamma-rays. As there is only minimal mixing with nickel-rich ejecta, this region cools primarily through hydrogen and helium line transitions. 

Table \ref{tab:tab1} shows the calculated properties of the B17 models. H$\alpha$ emission is strong and detectable for all of the models, regardless of whether companion is a MS, SG, or RG star.
To quantify this, we calculate the ratio of peak luminosity of \ha\ to that of [Fe III] $\lambda$ 4658  and find  $L_{\rm H\alpha}/L_{4658} > 0.8$ for all models. This prominent Balmer emission is inconsistent with observations of normal SNe~Ia and and so rules out progenitor scenarios with such high stripped masses. We find that the relative strength of the \ha\ emission is approximately constant over time, with $L_{\rm H\alpha}/L_{4658}$ decreasing by $ \lesssim 20\%$ from 200 to 500 days after explosion.

\begin{table}
\label{tab:tab1}
\centering
\caption{Derived properties of the B17 models. $M_{\rm st}$ refers to mass of material stripped from companion (taken from \citet{boehner2017}). $L_{\rm H\alpha}/L_{4658}$ is the ratio of peak flux of \ha\ to that of [Fe III] $\lambda$ 4658, not necessarily at line center. All luminosities reported are calculated at 200 days past explosion.}
\begin{tabular}{|c|c|c|c|}
\hline
\textbf{Model Name} & \textbf{$M_{\rm st}$} & \textbf{$L_{\rm H\alpha}$}&  \textbf{$L_{\rm H\alpha}/L_{4658}$} \\
&$(M_\odot)$& (10$^{39}$ erg s$^{-1}$) &\\
\hline
MS38 & 0.25 & 6.8 & 0.98 \\
\hline
MS7 & 0.37 & 9.3 & 1.43 \\
\hline
MS54 & 0.32 & 15.7 & 1.04 \\
\hline
MS63 & 0.24 & 7.0 & 1.23 \\
\hline
SG & 0.17 & 5.6 & 0.82 \\
\hline
RG319 & 0.28 & 4.5 & 0.84 \\
\hline
RG428 & 0.33 & 8.7 & 1.08 \\
\hline
\end{tabular}
\end{table}

\subsection{Reduced Mass Models}

The above models assumed a hydrogen companion in a semi-detached system and an idealized, spherically-symmetric W7-like explosion model. The total mass of stripped companion material, however, is found to lower if the SN explosion energy is decreased or the orbital separation distance is increased \citep{pakmor2008, pan2012, liu2012}. In addition, inhomogeneities in the  ejecta structure may disturb the flow and potentially affect the amount of stripped material \citet{hansen2007}.

Non-detections of \ha\ in observed SN~Ia nebular spectra clearly constrain the allowed mass of stripped material. However, converting the observed flux limit to quantitative mass constraints has been limited by the lack of detailed nebular modeling. Here we probe sensitivity of \ha\ flux to the amount of stripped material by gradually reducing the density in regions of MS38 model that contain at least 1\% hydrogen by mass. This simplified method should capture the basic effect of lowering the amount of stripped mass while retaining the characteristic asymmetry of the hydrodynamical interaction.

\begin{figure}[htbp]
\label{fig:ha_lum}
\includegraphics[width=0.5\textwidth]{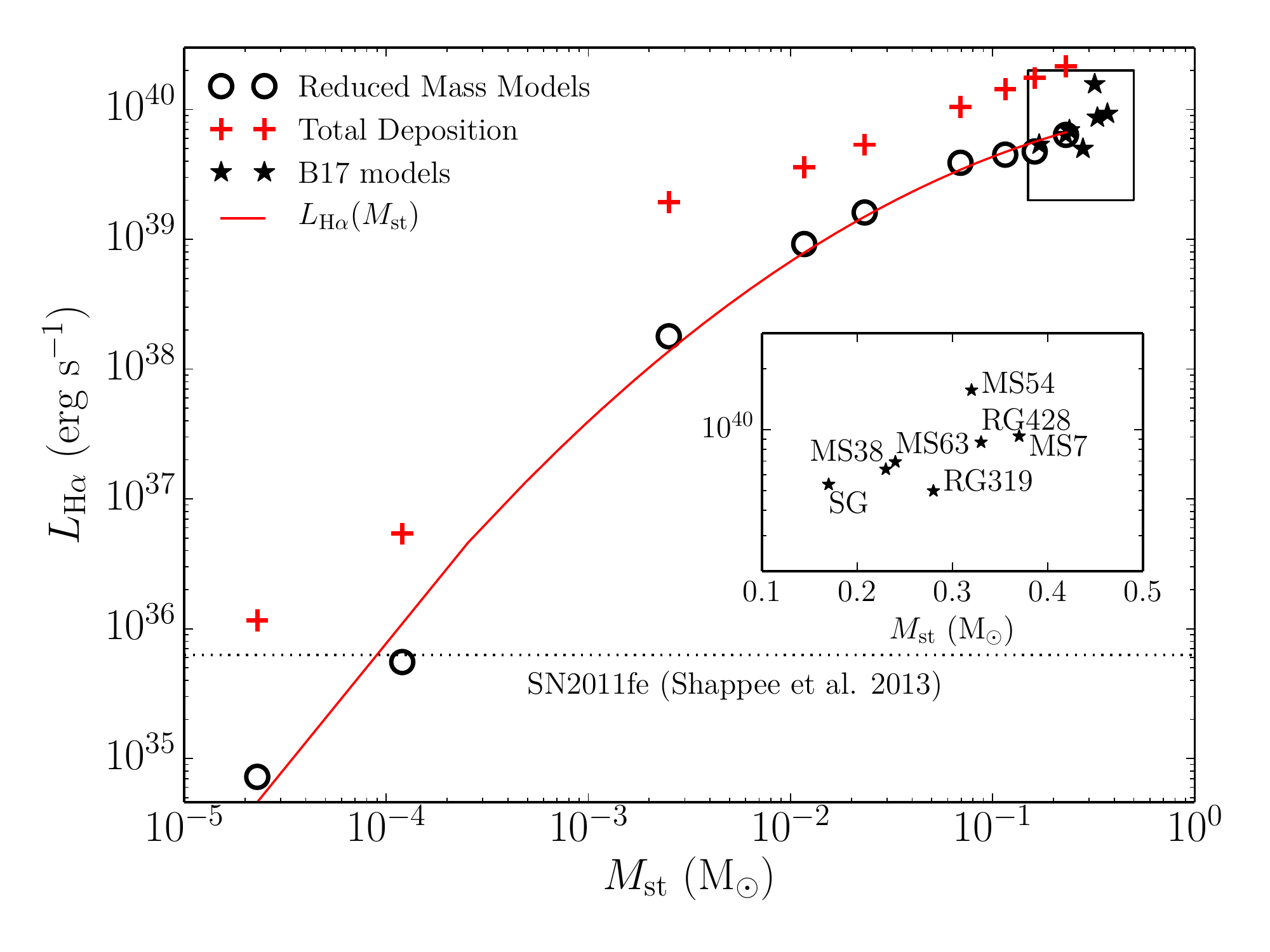}
\caption{\ha\ luminosity for various models as a function of stripped mass. Open circles denote the amount of stripped mass in a series of modified MS38 models, while filled stars label the original B17 models. Red plus symbols indicate the total energy deposited into the stripped material for each value of stripped mass. We found a close correlation between the amount of stripped mass in the original B17 models and their estimated \ha\ luminosities, which suggests that such dependence is a common characteristic of SN Ia in the SD scenario. The \ha\ luminosity upper limit for SN 2011fe \citep{shappee2013} is marked with a horizontal dashed line, scaled to the luminosity expected at 200 days past explosion (see text).
}
\end{figure}

Figure \ref{fig:ha_lum} shows the \ha\ model luminosity for a series of MS38 models with reduced density, along with the original set of B17 models. For the set of modified MS38 models we find that the \ha\ model luminosity approximately varies with the stripped mass as 

\begin{equation}
\log_{10}(L_{\rm H\alpha}) = -0.2 M_1^2 + 0.17 M_1 + 40.0,
\end{equation}
where $M_{\rm 1} = \log_{10}(M_{\rm st} / M_\odot)$ and $[L_{\rm H\alpha}] = \rm erg~s^{-1}$. We stress that this approximate relation holds only for or scaled MS38 models, and does not take into account how the stripped mass geometry may vary with separation distance, SN ejecta properties, or companion type.

We estimate that the \ha\ luminosity accounts for about 30\% of the total energy deposited in the stripped material for \sub{M}{st} $ > 0.01 M_\odot$, and scales with total deposition. At lower masses, hydrogen is more ionized and only accounts for about 10\% of total deposition energy. 

For reference, the \ha\ luminosity upper bound for SN 2011fe \citep{shappee2013} is marked with a horizontal dashed line in Figure \ref{fig:ha_lum}, scaled to account for the time difference between our models (200 days past explosion) and their observations (292 days past explosion). Since we find \ha\ luminosity to be a relatively constant fraction of bolometric luminosity, we can scale the \ha\ limit of SN 2011fe using its bolometric light curve \citep{mazzali2015}, in which there is roughly a factor of $\sim$4 decrease in luminosity between 200 and 292 days. Compared to our reduced mass models, the non-detection suggests a constraint on the stripped mass of $\lesssim 10^{-4}$ \Msun. This is a factor of about 5 lower than the constraint derived by \cite{shappee2013} based on parameterized 1D models. 

\subsection{Effect of viewing angle}
\label{sec:viewing}

\begin{figure}[htbp]
\label{fig:ms38_viewing}
\includegraphics[width=0.5\textwidth]{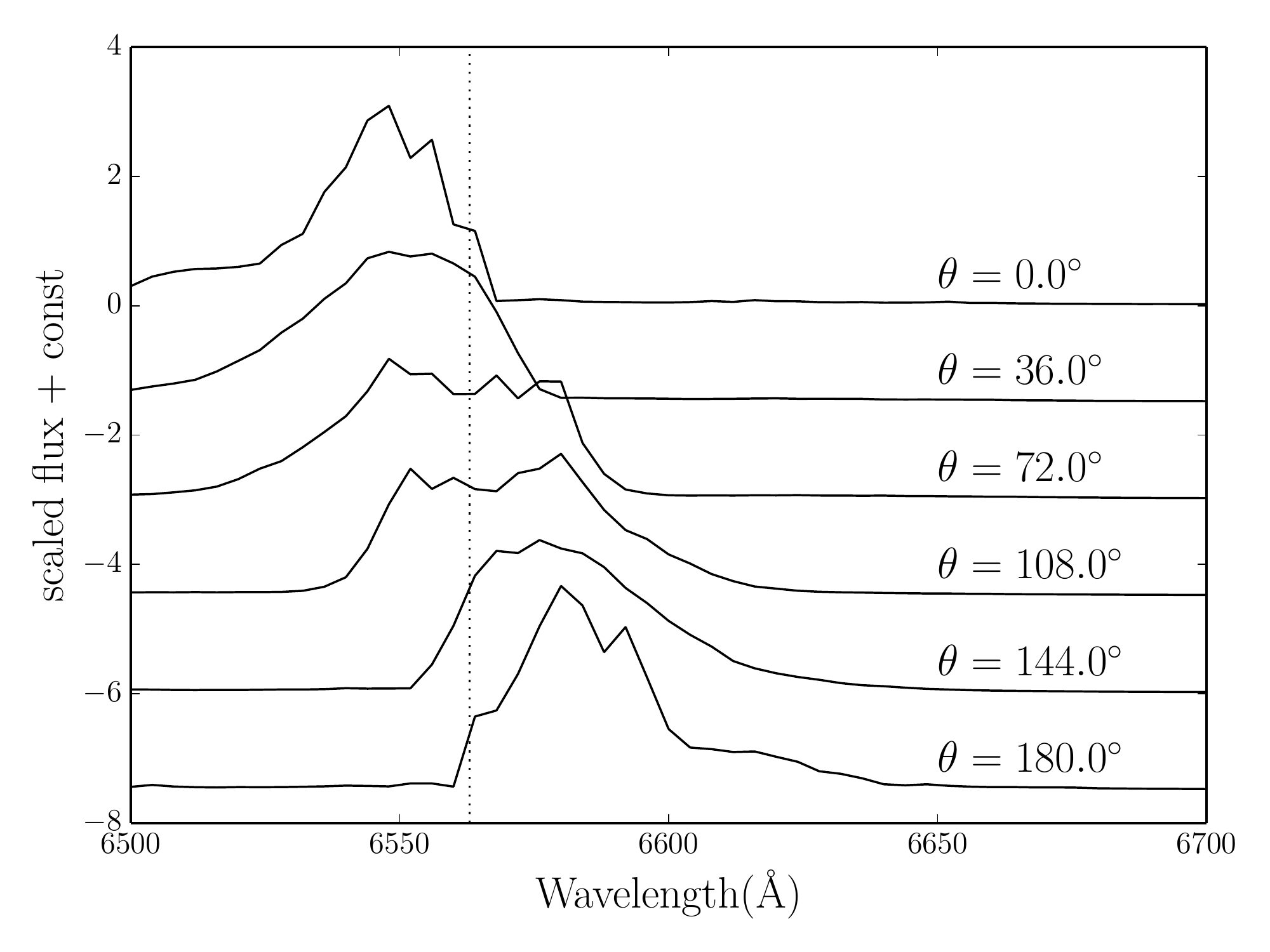}
\caption{Synthetic \ha\ profiles of the MS38 model obtained for multiple viewing angles. Line shifts are clear from comparison with the H$\alpha$ line center, shown with a dotted line. Note that the line profile is also affected by the viewing angle.}
\end{figure}

The asymmetry of companion-stripped material introduces a viewing angle dependence of the narrow emission line profiles. Figure \ref{fig:ms38_viewing} shows the synthetic nebular spectrum of the day 200 MS38 model as observed from a number of viewing angles. While the total integrated luminosity of \ha\ is independent of viewing angle in optically thin ejecta, the line profiles do depend on the orientation. At $\theta = 0^\circ$, the bulk of stripped material is moving toward the observer, resulting in a blue-shifted \ha\ peak. For $\theta = 180^\circ$, that region is moving away from the observer and the \ha\ peak is red-shifted. For intermediate angles, the feature is broader and has lower peak luminosity. In general, we expect that the \ha\ peak will be shifted by $\approx 10$~\AA\ from its rest wavelength due to orientation effects, which should be taken into account when trying to extract flux limits from observational data.

\subsection{Helium-only Models}

In addition to the hydrogen-rich systems considered by B17, helium star companion models have been proposed to explain SNe Ia \citep{iben1984, yoon2003, wang2009_companions, wang2009_he_star}. For such companions, hydrodynamic models of interaction with the ejecta predict that up to 0.06 \Msun\ of companion mass becomes unbound \citep{pan2010, pan2012, liu2013a_HeCompanions}. We explore such a scenario by replacing all hydrogen with helium in the solar-abundance stripped material of model M38. While such a modified model overestimates the amount of stripped mass by a factor of $\gtrsim 4$ and possibly inaccurately reflects the geometry of the stripped material, it does provide a first estimate of an \emph{upper bound} on the line emission from SD progenitors with helium companions.

\begin{figure*}[htbp]
\label{fig:fig3}
\includegraphics[width=\textwidth]{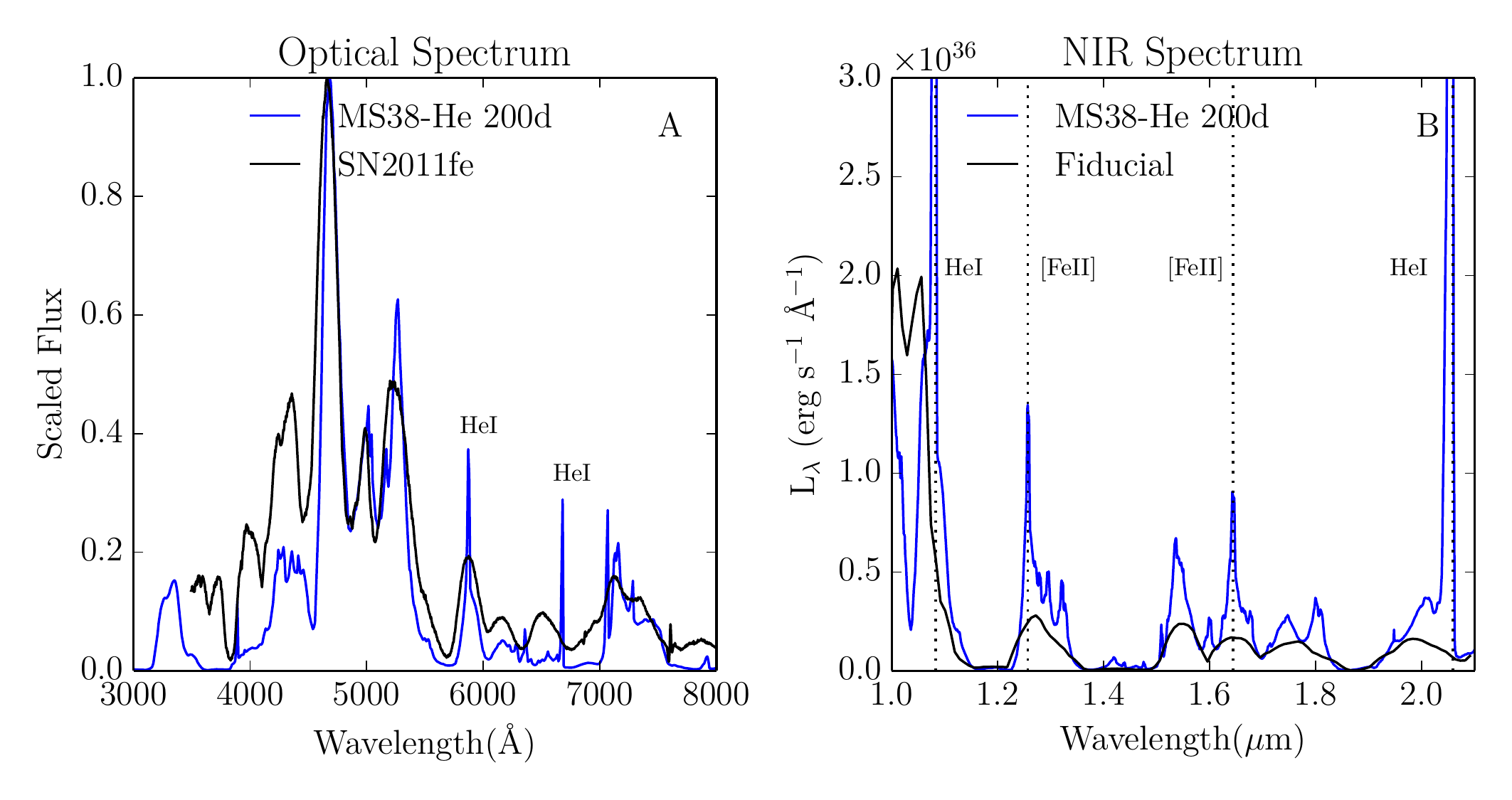}
\caption{Synthetic spectra of the MS38 model at 200 days past explosion with all hydrogen replaced by helium. Panel (A) shows the optical spectrum, which contains visible He I optical emission at He I $\lambda$5875, 6678. Panel (B) shows the synthetic NIR spectrum, which contains visible [Fe II] and HeI emission features. The fiducial model from \citet{botyanszki2017} is included in black for reference. Dotted lines identify the following transitions: He I 1.085$\mu$m, [Fe II] 12.257$\mu$m, [Fe II] 1.644 $\mu$m, He I 2.059$\mu$m. Line profiles are more peaked than in the fiducial model, and narrow He I emission dominates the NIR spectrum.}
\end{figure*}

Figure \ref{fig:fig3} shows the MS38 model with all stripped hydrogen replaced by helium. We find strong permitted helium emission at both optical and NIR wavelengths. In addition narrow emission lines from [CaII] and [FeII] lines are visible atop  broader components of line emission from the SN ejecta. The results suggest that even if the stripped helium mass is lower by a factor of $4$ or more, observations of high resolution nebular spectra can provide strong constraints of helium star SD progenitor models. 

\section{DISCUSSION}
\label{sec:discussion}

We have presented synthetic nebular spectra of multidimensional companion-ejecta interaction models in binary systems representing classic SD progenitors of SNe~Ia.  We found that the late-time \ha\ emission in SN~Ia models with hydrogen-rich companions (regardless whether MS, SG, or RG type), which results in $\gtrsim$ 0.1 \Msun\ of unbound solar-abundance ejecta, is strong compared to the limits derived from observed spectra. The \ha\ emission in our realistic models is even stronger than that found in previous parameterized 1D studies. We therefore disfavor semi-detached binary systems with hydrogen-rich companions as viable SN~Ia progenitors.

By artificially reducing the mass of the stripped region, we estimated that even an order of magnitude less of stripped, hydrogen-rich material (i.e., $\approx$ 0.01 \Msun) would still be observationally detectable. Consequently, we place a strong limit of $M_{\rm st} \lesssim 1\times 10^{-4}$ \Msun\ on the  hydrogen mass for SN2011fe \citep{shappee2013}, assuming that our models are representative of the  geometry of stripped material. This constraint is a factor of 5 stronger limit than that derived by \cite{shappee2013} based on parameterized 1D models.

For SD scenarios to be consistent with the non-detection of \ha\ in normal SN Ia requires much smaller stripped hydrogen masses. This may be possible if the ratio of the binary separation distance, $a$, to the companion star radius $R$ is larger than the value $a/R \approx 3$ adopted for the Roche lobe-filling orbital geometry of the B17 models. For larger $a/R$, the solid angle subtended by the companion is smaller, and a smaller fraction of ejecta is intercepted. Based on the scalings of \citet{liu2012}, the stripped mass may be consistent with $\lesssim 10^{-4}$~\Msun\ if  $a/R \gtrsim 20$. However, a supplementary analysis conducted using the method of \citet{boehner2017} suggests that the dependence of stripped mass on binary separation might be more shallow than previously reported, leading to stripped masses above 0.1 \Msun\ even as binary separation increases by a factor of 8. More work is needed to clarify the relationship between late-time \ha\ luminosity and binary separation. Furthermore, certain scenarios, such as the spin-up/spin-down models of \citet{distefano2011, justham2011}, posit a delay between accretion and explosion during which the companion can shrink by orders of magnitude in radius. 

For SD scenarios involving a non-degenerate helium star companion, the interaction with the ejecta results in a lower amount of stripped mass, $\approx$ 0.06 \Msun\ \citep{pan2010, pan2012, liu2013a_HeCompanions}. Our rather speculative MS38-based helium-only model still showed significant He I spectral features. As our modified model contained about 4 times more mass of stripped helium than found in actual simulations, we consider our helium emission estimates as an upper bound to the emission detectable from more realistic models.

Future work should consider a broader range of hydrodynamical models, as the B17 sample exclusively considers SD, Roche-lobe filling, hydrogen-rich companions. Both the orbital geometry (wider systems) and companion type (helium star) should be considered in more detail. An idealized, spherically-symmetric W7 explosion model provides for smooth, steady flow of ejecta around the companion. However, realistic SN Ia explosion models will likely produce ejecta characterized by density inhomogeneities that will perturb the interaction region potentially enhancing the stripping process \citep{hansen2007}. Finally, the limitations in input atomic data and the approximations in the treatment of atomic processes and radiative transfer effects discussed in \citet{botyanszki2017} influence nebular spectral modeling and should be addressed in future studies.

\acknowledgments{
  This material is based upon work supported by the National Science Foundation Graduate Research Fellowship Program under Grant No. DGE 1106400. DK is supported in part by a Department of Energy Office of Nuclear Physics Early Career Award, and by the Director, Office of Energy Research, Office of High Energy and Nuclear Physics, Divisions of Nuclear Physics, of the U.S. Department of Energy under Contract No. DE-AC02-05CH11231. This research used resources of the National Energy Research Scientific Computing Center, a DOE Office of Science User Facility supported by the Office of Science of the U.S. Department of Energy under Contract No. DE-AC02-05CH11231.
}

\end{document}